\documentclass[manuscript]{raa}
\usepackage{graphicx,times}             
\usepackage{natbib}
\usepackage{amsfonts}
\usepackage{txfonts}

\newcommand{\Rmnum}[1]{\expandafter\@slowromancap\romannumeral #1@}
\begin{document}

   \title{Detailed study of B037 based on {\sl HST} images
}
   \volnopage{Vol.0 (200x) No.0, 000--000}      
   \setcounter{page}{1}           

   \author{Jun Ma \mailto{}
      \inst{1,2},
      }

   \institute{$^{1}$National Astronomical Observatories, Chinese
Academy of Sciences, Beijing, 100012, P. R. China\\
$^{2}$Key Laboratory of Optical Astronomy, National Astronomical Observatories, Chinese Academy of
Sciences, Beijing, 100012, China\\
             \email{majun@nao.cas.cn}
          }

   \date{Received~~2001 month day; accepted~~2001~~month day}
   \authorrunning{Wu et al.}            
   \titlerunning{Multiwavelength study of low-luminosity 6.7-GHz methanol masers}  

\abstract{B037 is of interest because it is both the most luminous and the most highly reddened
cluster known in M31. Images of deep observations and of highly spatial resolutions with the
Advanced Camera for Surveys on the {\sl Hubble Space Telescope (HST)} firstly show that this
cluster is crossed by a dust lane. Photometric data in the F606W and F814W filters obtained in
this paper provide that, colors of ($\rm {F606W-F814W}$) in the dust lane are redder $\sim 0.4$
mags than ones in the other regions of B037. The {\sl HST} images show that, this dust lane seems
to be contained in B037, not from the M31 disk or the Milky Way. As we know, the formation of dust
requires gas with a rather high metallicity. However, B037 has a low metallicity to be $\rm
[Fe/H]=-1.07\pm 0.20$. So, it seems improbable that the observed dust lane is physically
associated with B037. It is clear that the origin of this dust lane is worthy of future study. In
addition, based on these images, we present the precise variation of ellipticity and position
angle, and of surface brightness profile, and determine the structural parameters of B037 by
fitting a single-mass isotropic King model. In the F606W filter, we derive the best-fitting scale
radius, $r_0=0.56\pm0.02\arcsec~(=2.16\pm0.08~\rm{pc})$, a tidal radius,
$r_t=8.6\pm0.4\arcsec~(=33.1\pm1.5~\rm{pc})$, and a concentration index $c=\log
(r_t/r_0)=1.19\pm0.02$. In the F814W filter, we derive
$r_0=0.56\pm0.01\arcsec~(=2.16\pm0.04~\rm{pc})$, $r_t=8.9\pm0.3\arcsec~(=34.3\pm1.2~\rm{pc})$, and
$c=\log (r_t/r_0)=1.20\pm0.01$. The extinction-corrected central surface brightness is
$\mu_0=13.53\pm 0.03~{\rm mag~arcsec^{-2}}$ in the F606W filter, and $12.85\pm 0.03~{\rm
mag~arcsec^{-2}}$ in the F814W filter, respectively. We also calculate the half-light radius, at
$r_h=1.05\pm0.03\arcsec(=4.04\pm0.12~\rm{pc})$ in the F606W filter and
$r_h=1.07\pm0.01\arcsec(=4.12\pm0.04~\rm{pc})$ in the F814W filter, respectively. In addition, we
derived the whole magnitudes of B037 in $V$ and $I$ bands by transforming the magnitudes from the
ACS system to the standard system, which are in very agreement with the previous ground-based
broad-band photometry.
   \keywords{galaxies: evolution -- galaxies: individual (M31) --
globular cluster: individual (B037)}
   }
   \authorrunning{Jun Ma}            
   \titlerunning{Detailed study of B327 based on {\sl HST} images}  
   \maketitle

\section{Introduction}

Globular clusters (GCs) are effective laboratories for studying stellar evolution and stellar
dynamics, and they are ancient building blocks of galaxies which can help us to understand the
formation and evolution of their parent galaxies. In addition, GCs exhibit surprisingly uniform
properties, suggesting a common formation mechanism.

The closest other populous GC system beyond the halo of our Galaxy is that of M31. The study of
M31 has been and continues to be a keystone of extragalactic astronomy \citep{bh00}, and the study
of GCs in M31 can be traced back to \citet{hubble32}. M31 GC B327 (B for `Baade') or Bo37 (Bo for
`Bologna', see Battistini 1987), which, in the nomenclature introduced by \citet{huchra91} is
referred to as B037, a designation from the Revised Bologna Catalogue (RBC) of M31 GCs and
candidates (Galleti et al. 2004, 2006, 2007), which is the main catalog used in studies of M31
GCs. The extremely red color of B037 was firstly noted by \citet{kronmay60}, who suggested that
this cluster must be highly reddened. Two years later, \citet{vete62a} determined magnitudes of
257 M31 GC candidates including B037 in the $UBV$ photometric system, and then \citet{vete62b}
studied the intrinsic colors of M31 GCs, and found that B037 was the most highly reddened with
$E(B-V)=1.28$ in his sample of M31 GC candidates based on the photometric catalog of
\citet{vete62a}. With low-resolution spectroscopy, \citet{cram85} also found that B037 is the most
highly reddened GC candidate in M31 to have $E(B-V)=1.48$. Based on a large database of multicolor
photometry, \citet{bh00} determined the reddening value for each individual M31 GC including B037
using the correlations between optical and infrared colors and metallicity by defining various
``reddening-free'' parameters, and the reddening value of B037 is $E(B-V)=1.38\pm0.02$ (which is
kindly given us by P. Barmby). Again, \citet{bk02} derived the reddening value for this cluster to
be $E(B-V)=1.30\pm0.04$, using the spectroscopic metallicity to predict the intrinsic colors.
\citet{ma06b} also determine the reddening of B037 by comparing independently obtained multicolor
photometry with theoretical stellar population synthesis models to be $E(B-V)=1.360\pm0.013$,
which is in good agreement with the other results. Following the methods of \citet{bh00},
\citet{fan08} (re-)determined reddening values for 443 clusters and cluster candidates including
B037, and the redding value of B037 obtained by \citet{fan08} is $E(B-V)=1.21\pm0.03$, which is a
little smaller than the previous determinations.

The brightest GCs in M31 are more luminous than the most brightest Galactic cluster, $\omega$
Centauri. Among these are B037 \citep{Sidney68} and G1 \citep[see details from][]{bk02}. These two
clusters are both considered as the possible remnant core of a former dwarf galaxy which lost most
of its envelope through tidal interactions with M31 \citep{meylan97,meylan01,mackey05,ma06a,ma07}.

In this paper, we will present the photometric data of B037 using its deep images obtained with
the Advanced Camera for Survey (ACS) on the {\sl HST}. The deep images of highly spatial
resolutions showed that this cluster is crossed by a dust lane. Our results provide that colors of
F814W$-$F606W in the dust lane are redder $\sim 0.4$ mags than ones of the other regions. In
addition, we studied structures of B037 in detail based on these images.

\section{Observations and Data Reduction}

\subsection{{\sl HST} images of B037}

We searched the {\sl HST} archive and found B037 to have been observed with the ACS-Wide Field
Channel (WFC) in the F606W and the F814W filters, which were observed on 2004 August 2 and on 2004
July 4, respectively. The exposure time is 2370.0 seconds for both bands. The {\sl HST} ACS-WFC
resolution is $0.05\arcsec$ per pixel. The images in F606W and F814W both show that B037 is
crossed by a dust lane. Fig. 1 clearly shows the dust lane, which crosses B037. If the dust lane
is true, its color should be different from ones of the other regions.

If not otherwise stated, the magnitudes are always on the VEGAMAG scale as defined by
\citet{sirianni05}. The relevant zero-point for this system is 26.398 and 25.501 for WFC F606W and
WFC F814W, respectively. A distance to M31 of 780 kpc ($1{\arcsec}$ subtends 3.85 pc) is adopted
in this paper.

\begin{figure}
\begin{center}
\includegraphics[width=0.75\textwidth]{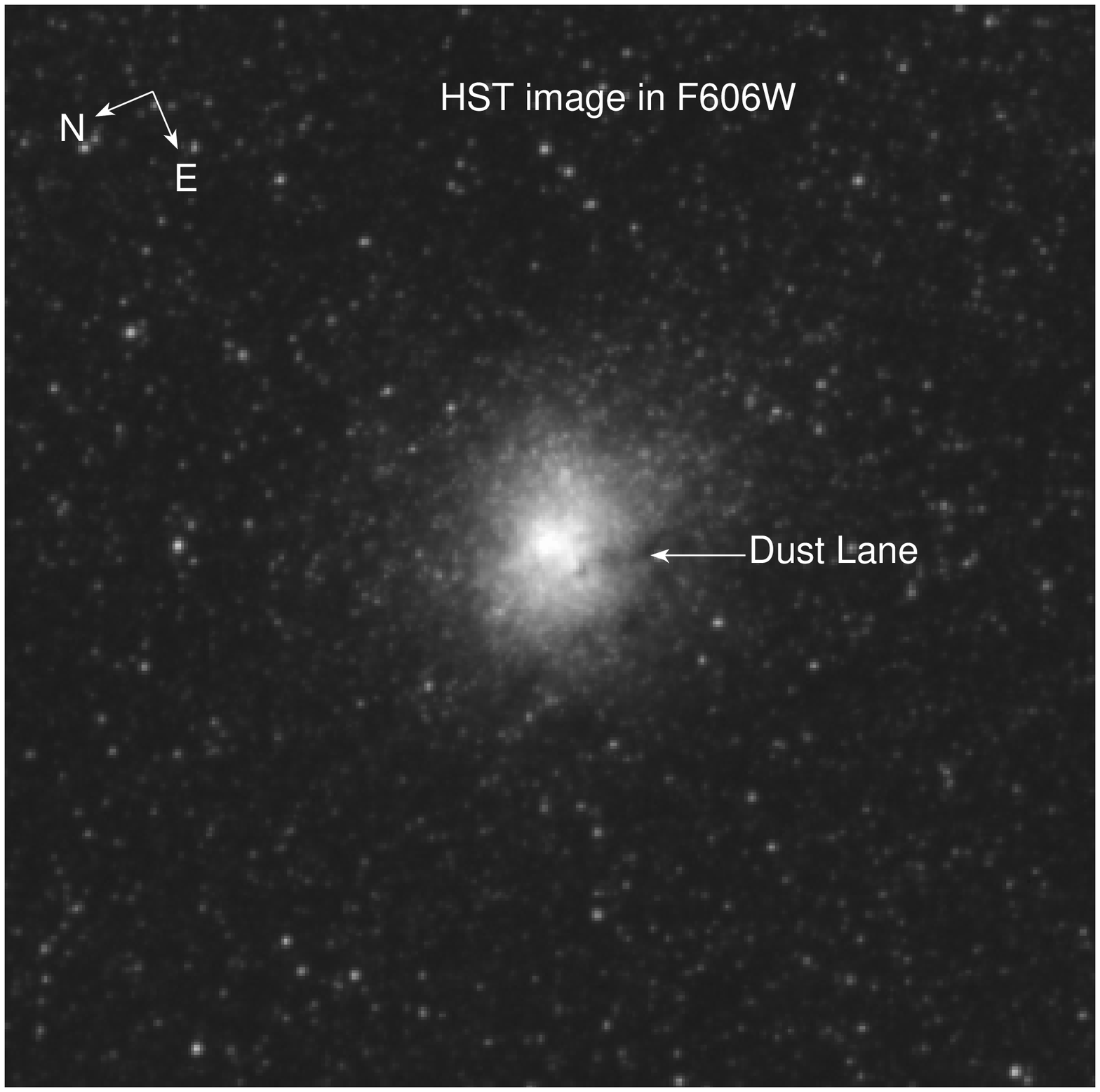}\\
\includegraphics[width=0.75\textwidth]{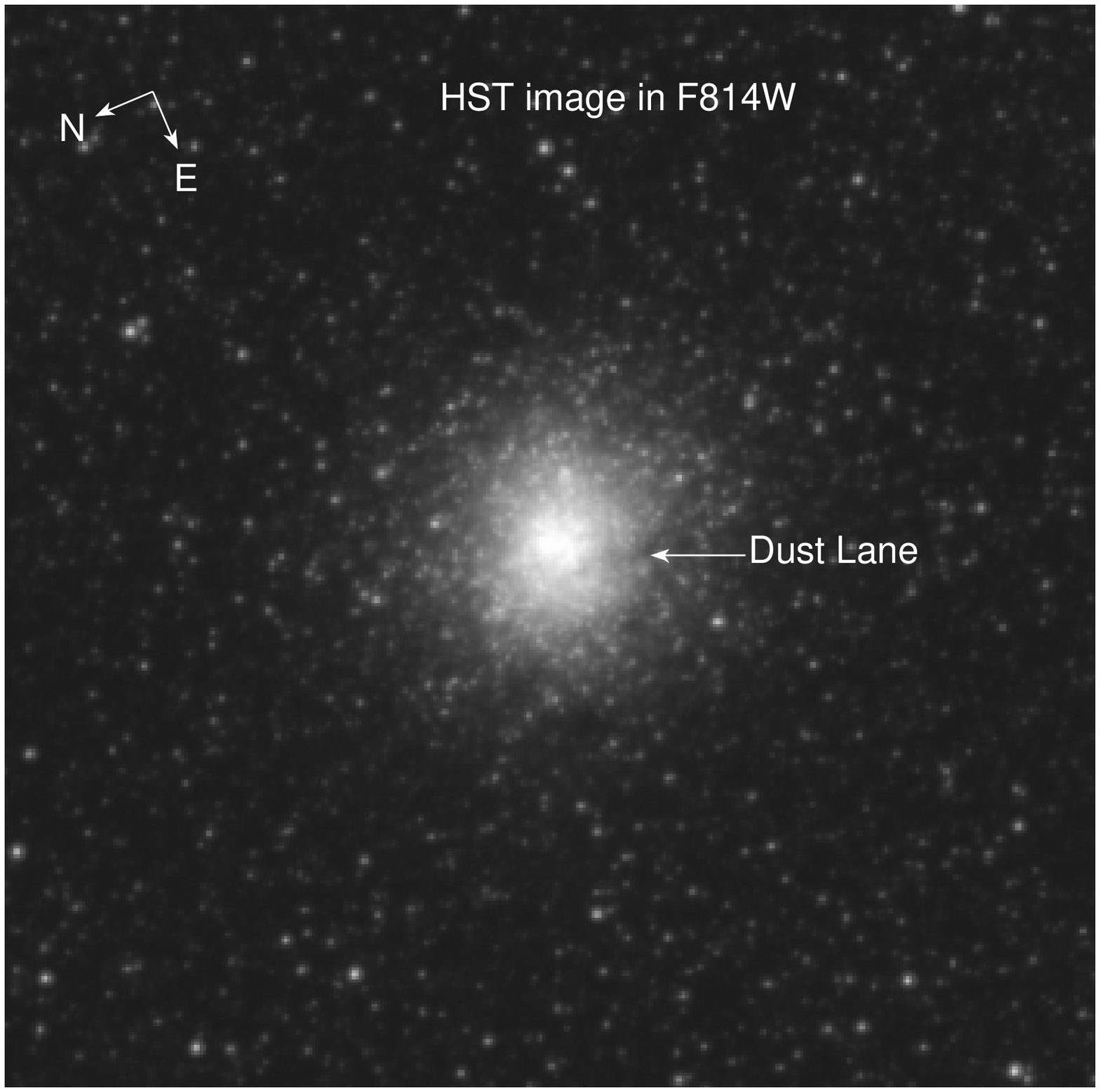}
\caption{The images of GC B037 observed in the F606W and F814W filters of ACS/{\sl HST}. The
images clearly show that the cluster is crossed by a dust lane. The image size is
$17.5\arcsec\times17.5\arcsec$ for each panel.} \label{fig1}
\end{center}
\end{figure}

\subsection{Color difference between the dust lane and the other regions}

In order to study whether the color difference between the dust lane and the other regions in B037
exists, we select nine points, three of which (No. 7, 8 and 9) are located in the dust lane, the
other six are randomly located in the other regions (see Figure 2). For each sample point, the
PHOT routine in DAOPHOT \citep{stet87} is used to obtain magnitude. We adopt an aperture of a
diameter of 4 pixels. The photometric data for these nine sample points are given in Table 1, in
conjunction with the $1\sigma$ magnitude uncertainties from {\sc daophot}. Column 4 gives the
color of ($\rm {F606W-F814W}$). From Table 1, we can see that colors of ($\rm {F606W-F814W}$) in
the dust lane are redder $\sim 0.4$ mags than ones of the other regions.

\begin{figure}
\begin{center}
\includegraphics[height=140mm,angle=0]{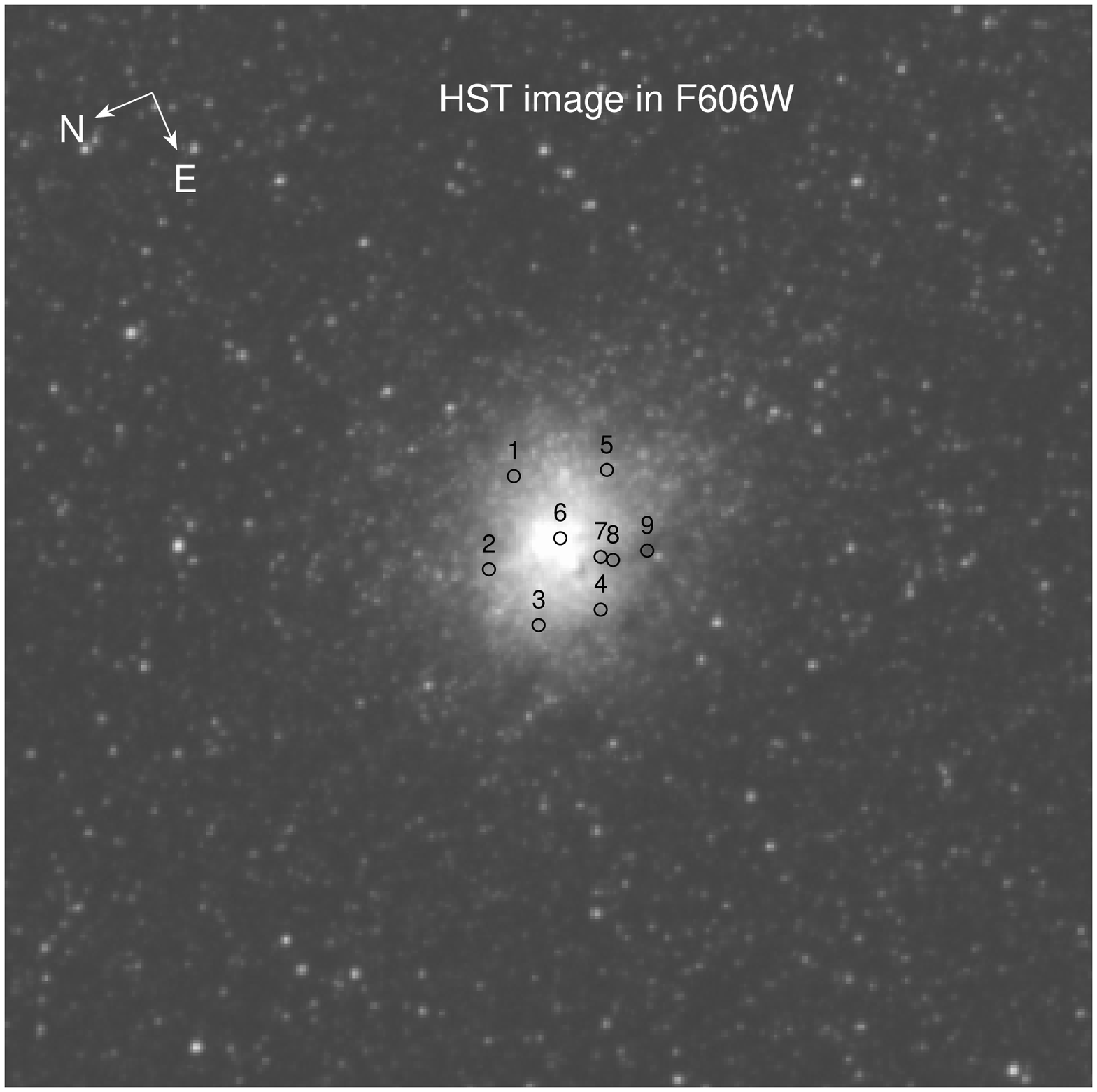}%
\caption{\label{Fig:maps}The sample positions of photometry (black circles) are showed in the
image of GC B037 observed in the F606W filter of ACS/{\sl HST}. An aperture of radii of 2 pixels
is adopted for photometry. The image size is $17.5\arcsec\times17.5\arcsec$.}
\end{center}
\end{figure}

\begin{table}
\begin{center}
\renewcommand\arraystretch{1}
\caption{\label{Tab:assocation} Photometric data for B037}
\begin{tabular}{cccc}
\hline \hline
Source & F606W  & F814W  & F606W$-$F814W \\
No.    & (mag) & (mag) & (mag) \\
\hline
1 & $ 23.29 \pm 0.26$ & $ 21.35 \pm 0.16$ & $ 1.94 $ \\
2 & $ 23.03 \pm 0.23$ & $ 21.14 \pm 0.15$ & $ 1.89 $ \\
3 & $ 22.95 \pm 0.22$ & $ 21.00 \pm 0.14$ & $ 1.95 $ \\
4 & $ 23.22 \pm 0.25$ & $ 21.25 \pm 0.15$ & $ 1.97 $ \\
5 & $ 23.17 \pm 0.25$ & $ 21.36 \pm 0.16$ & $ 1.81 $ \\
6 & $ 21.15 \pm 0.10$ & $ 19.10 \pm 0.06$ & $ 2.05 $ \\
7 & $ 22.93 \pm 0.22$ & $ 20.53 \pm 0.11$ & $ 2.40 $ \\
8 & $ 23.41 \pm 0.28$ & $ 20.94 \pm 0.13$ & $ 2.47 $ \\
9 & $ 24.66 \pm 0.49$ & $ 22.31 \pm 0.25$ & $ 2.35 $ \\
\hline \hline
\end{tabular}
\end{center}
\end{table}

\subsection{Surface brightness profiles}

We used the {\sc iraf} task {\sc ellipse} to obtain F606W and F814W surface brightness profiles
for B037. B037 center position was fixed at a value derived by object locator of {\sc ellipse}
task, however an initial center position was determined by centroiding. Elliptical isophotes were
fitted to the data, with no sigma clipping. We ran two passes of {\sc ellipse} task, the first
pass was run in the usual way, with ellipticity and position angle allowed to vary with the
isophote semimajor axis. In the second pass, surface brightness profiles on fixed,
zero-ellipticity isophotes were measured, since we choose to fit circular models for the intrinsic
cluster structure and the point spread function (PSF) as \citet{barmby07} did (see \S 2.4 for
details). The background value was derived as the mean of a region of $100\times100$ pixels in
``empty'' areas far away from the cluster.

\subsubsection{Ellipticity and position angle}

Tables 2 and 3 give the ellipticity, $\epsilon=1-b/a$, and the position angle (P.A.) as a function
of the semi-major axis length, $a$, from the center of annulus in the F606W and F814W filter
bands, respectively. These observables have also been plotted in Figures 3 and 4, respectively;
the errors were generated by the {\sc iraf} task {\sc ellipse}, in which the ellipticity errors
are obtained from the internal errors in the harmonic fit, after removal of the first and second
fitted harmonics. From Table 3, and Figs. 3 and 4, we can see that, the values of ellipticity and
position angle cannot be obtained within $0.1448\arcsec$ in the F814W filter because of very high
ellipticity ($>1.0$). \citet{ma06a} analyzed the same F606W image of B037 used here, fitting a
\citet{king62} model to a surface brightness profile made from a PSF-deconvolved image. They also
plotted the distributions of ellipticity and the position angle as a function of the semi-major
axis length. Comparison of Fig. 2 of \citet{ma06a} and Figs. 3 and 4 shows that, the general trend
of the cluster's ellipticity as a function of semimajor axis radius is similar between
\citet{ma06a} and the present paper. The comparison also shows that uncertainties in the exact
value of the PA are only of secondary importance for the general trend in ellipticity observed,
given that the PA determination between \citet{ma06a} and the present paper differs somewhat
greatly. There are a number of possible reasons for the offsets in PA observed between these two
studies. The main reason is that, \citet{ma06a} used the PSF-deconvolved image. Other reasons
include those related to the positions of the centering of isophotes and the different geometrical
parameters set when fitting. In addition, Fig. 3 shows that the ellipticity varies significantly
with position along the semimajor axis radius, especially smaller than $0.5\arcsec$. In the F814W
filter band, the ellipticity is larger than 1.0 along the semimajor axis radius smaller than
$0.1448\arcsec$.

\begin{table}
\begin{center}
\renewcommand\arraystretch{1}
\caption{B037: Ellipticity, $\epsilon$, and position angle (P.A.) as a function of the semimajor
axis, $a$, in the F606W filter of {\sl HST} ACS-WFC}
\begin{tabular}{ccc|ccc}
\hline\hline
   $a$    &  $\epsilon$  &   P.A.   &   $a$    &  $\epsilon$  &   P.A. \\
 (arcsec) &              &  (deg)   & (arcsec) &      &   (deg)        \\
\hline
 0.0260  & $0.638   \pm0.228   $ & $92.9    \pm15.5    $ & 0.3757  & $0.177   \pm0.031   $ & $69.8    \pm5.6     $ \\
 0.0287  & $0.638   \pm0.229   $ & $93.2    \pm15.6    $ & 0.4132  & $0.151   \pm0.029   $ & $64.7    \pm5.9     $ \\
 0.0315  & $0.639   \pm0.230   $ & $93.4    \pm15.7    $ & 0.4545  & $0.090   \pm0.027   $ & $60.3    \pm9.1     $ \\
 0.0347  & $0.640   \pm0.232   $ & $93.7    \pm15.8    $ & 0.5000  & $0.005   \pm0.025   $ & $172.6   \pm30.0    $ \\
 0.0381  & $0.642   \pm0.233   $ & $94.0    \pm15.9    $ & 0.5500  & $0.060   \pm0.020   $ & $155.5   \pm9.8     $ \\
 0.0420  & $0.643   \pm0.235   $ & $94.4    \pm16.0    $ & 0.6050  & $0.117   \pm0.015   $ & $156.3   \pm3.9     $ \\
 0.0461  & $0.645   \pm0.236   $ & $94.7    \pm16.0    $ & 0.6655  & $0.174   \pm0.012   $ & $157.2   \pm2.2     $ \\
 0.0508  & $0.647   \pm0.182   $ & $95.2    \pm12.3    $ & 0.7321  & $0.233   \pm0.011   $ & $157.2   \pm1.5     $ \\
 0.0558  & $0.599   \pm0.159   $ & $96.8    \pm11.2    $ & 0.8053  & $0.278   \pm0.011   $ & $159.1   \pm1.3     $ \\
 0.0614  & $0.546   \pm0.142   $ & $98.5    \pm10.6    $ & 0.8858  & $0.322   \pm0.011   $ & $160.8   \pm1.1     $ \\
 0.0676  & $0.503   \pm0.127   $ & $100.2   \pm10.0    $ & 0.9744  & $0.358   \pm0.012   $ & $162.1   \pm1.2     $ \\
 0.0743  & $0.458   \pm0.099   $ & $102.2   \pm8.3     $ & 1.0718  & $0.380   \pm0.021   $ & $164.5   \pm2.1     $ \\
 0.0818  & $0.400   \pm0.059   $ & $104.3   \pm5.5     $ & 1.1790  & $0.367   \pm0.022   $ & $168.0   \pm2.2     $ \\
 0.0899  & $0.410   \pm0.050   $ & $101.0   \pm4.6     $ & 1.2969  & $0.343   \pm0.025   $ & $169.2   \pm2.7     $ \\
 0.0989  & $0.428   \pm0.044   $ & $98.7    \pm3.9     $ & 1.4266  & $0.319   \pm0.025   $ & $166.8   \pm2.8     $ \\
 0.1088  & $0.437   \pm0.046   $ & $97.6    \pm3.9     $ & 1.5692  & $0.252   \pm0.022   $ & $165.1   \pm3.0     $ \\
 0.1197  & $0.428   \pm0.028   $ & $96.6    \pm2.5     $ & 1.7261  & $0.239   \pm0.021   $ & $165.7   \pm2.9     $ \\
 0.1317  & $0.410   \pm0.027   $ & $96.5    \pm2.5     $ & 1.8987  & $0.211   \pm0.026   $ & $163.6   \pm4.0     $ \\
 0.1448  & $0.400   \pm0.031   $ & $96.3    \pm3.0     $ & 2.0886  & $0.201   \pm0.029   $ & $152.8   \pm4.6     $ \\
 0.1593  & $0.364   \pm0.023   $ & $95.1    \pm2.3     $ & 2.2975  & $0.188   \pm0.037   $ & $150.1   \pm6.3     $ \\
 0.1752  & $0.352   \pm0.027   $ & $94.8    \pm2.8     $ & 2.5272  & $0.182   \pm0.033   $ & $149.9   \pm5.8     $ \\
 0.1928  & $0.337   \pm0.027   $ & $93.2    \pm2.9     $ & 2.7800  & $0.180   \pm0.034   $ & $145.6   \pm6.0     $ \\
 0.2120  & $0.311   \pm0.027   $ & $92.5    \pm3.0     $ & 3.0580  & $0.191   \pm0.031   $ & $137.5   \pm5.2     $ \\
 0.2333  & $0.287   \pm0.026   $ & $90.6    \pm3.1     $ & 3.3638  & $0.143   \pm0.034   $ & $125.4   \pm7.2     $ \\
 0.2566  & $0.258   \pm0.026   $ & $88.6    \pm3.3     $ & 3.7001  & $0.180   \pm0.041   $ & $121.1   \pm7.1     $ \\
 0.2822  & $0.233   \pm0.027   $ & $85.4    \pm3.8     $ & 4.0701  & $0.257   \pm0.033   $ & $121.6   \pm4.1     $ \\
 0.3105  & $0.207   \pm0.029   $ & $81.2    \pm4.5     $ & 4.4772  & $0.233   \pm0.048   $ & $121.6   \pm6.5     $ \\
 0.3415  & $0.189   \pm0.030   $ & $75.3    \pm5.0     $ & 4.9249  & $0.237   \pm0.063   $ & $116.1   \pm8.5     $ \\
\hline \hline
\end{tabular}
\end{center}
\end{table}

\begin{table}
\begin{center}
\renewcommand\arraystretch{1}
\caption{B037: Ellipticity, $\epsilon$, and position angle (P.A.) as a function of the semimajor
axis, $a$, in the F814W filter of {\sl HST} ACS-WFC}
\begin{tabular}{ccc|ccc}
\hline\hline
   $a$    &  $\epsilon$  &   P.A.   &   $a$    &  $\epsilon$  &   P.A. \\
 (arcsec) &              &  (deg)   & (arcsec) &      &   (deg)        \\
\hline
 0.0260  &                       &                       & 0.4132  & $0.031   \pm0.030   $ & $77.8    \pm28.3    $ \\
 0.0287  &                       &                       & 0.4545  & $0.031   \pm0.029   $ & $20.2    \pm27.5    $ \\
 0.0315  &                       &                       & 0.5000  & $0.044   \pm0.027   $ & $161.2   \pm17.6    $ \\
 0.0347  &                       &                       & 0.5500  & $0.084   \pm0.023   $ & $150.3   \pm8.3     $ \\
 0.0381  &                       &                       & 0.6050  & $0.127   \pm0.021   $ & $150.3   \pm5.0     $ \\
 0.0420  &                       &                       & 0.6655  & $0.175   \pm0.019   $ & $153.6   \pm3.4     $ \\
 0.0461  &                       &                       & 0.7321  & $0.220   \pm0.016   $ & $157.4   \pm2.3     $ \\
 0.0508  &                       &                       & 0.8053  & $0.247   \pm0.013   $ & $161.8   \pm1.7     $ \\
 0.0558  &                       &                       & 0.8858  & $0.251   \pm0.014   $ & $167.8   \pm1.8     $ \\
 0.0614  &                       &                       & 0.9744  & $0.263   \pm0.017   $ & $170.0   \pm2.1     $ \\
 0.0676  &                       &                       & 1.0718  & $0.293   \pm0.034   $ & $170.8   \pm4.0     $ \\
 0.0743  &                       &                       & 1.1790  & $0.297   \pm0.035   $ & $172.5   \pm4.1     $ \\
 0.0818  &                       &                       & 1.2969  & $0.230   \pm0.028   $ & $171.5   \pm4.0     $ \\
 0.0899  &                       &                       & 1.4266  & $0.216   \pm0.025   $ & $166.7   \pm3.7     $ \\
 0.0989  &                       &                       & 1.5692  & $0.198   \pm0.031   $ & $165.5   \pm5.1     $ \\
 0.1088  &                       &                       & 1.7261  & $0.198   \pm0.025   $ & $169.5   \pm4.0     $ \\
 0.1197  &                       &                       & 1.8987  & $0.188   \pm0.029   $ & $167.5   \pm5.0     $ \\
 0.1317  &                       &                       & 2.0886  & $0.139   \pm0.031   $ & $166.9   \pm6.9     $ \\
 0.1448  &                       &                       & 2.2975  & $0.117   \pm0.031   $ & $117.8   \pm8.0     $ \\
 0.1593  & $0.908   \pm0.117   $ & $89.9    \pm6.9     $ & 2.5272  & $0.100   \pm0.034   $ & $148.1   \pm10.2    $ \\
 0.1752  & $0.878   \pm0.026   $ & $90.9    \pm1.5     $ & 2.7800  & $0.118   \pm0.051   $ & $141.0   \pm13.3    $ \\
 0.1928  & $0.827   \pm0.151   $ & $90.8    \pm9.6     $ & 3.0580  & $0.094   \pm0.043   $ & $115.8   \pm13.7    $ \\
 0.2120  & $0.749   \pm0.034   $ & $90.3    \pm2.3     $ & 3.3638  & $0.094   \pm0.035   $ & $132.7   \pm11.1    $ \\
 0.2333  & $0.731   \pm0.037   $ & $89.4    \pm2.6     $ & 3.7001  & $0.103   \pm0.056   $ & $121.3   \pm16.4    $ \\
 0.2566  & $0.695   \pm0.041   $ & $87.1    \pm2.9     $ & 4.0701  & $0.127   \pm0.061   $ & $120.1   \pm14.6    $ \\
 0.2822  & $0.624   \pm0.031   $ & $84.4    \pm2.3     $ & 4.4772  & $0.162   \pm0.031   $ & $115.7   \pm5.9     $ \\
 0.3105  & $0.546   \pm0.035   $ & $80.4    \pm2.6     $ & 4.9249  & $0.091   \pm0.054   $ & $131.4   \pm17.7    $ \\
 0.3415  & $0.401   \pm0.035   $ & $72.4    \pm3.2     $ & 5.4174  & $0.150   \pm0.065   $ & $135.5   \pm13.3    $ \\
 0.3757  & $0.258   \pm0.029   $ & $63.4    \pm3.8     $ & 5.9591  & $0.188   \pm0.044   $ & $161.8   \pm7.4     $ \\
\hline \hline
\end{tabular}
\end{center}
\end{table}

\begin{figure}
\centering
\includegraphics[height=150mm,angle=-90]{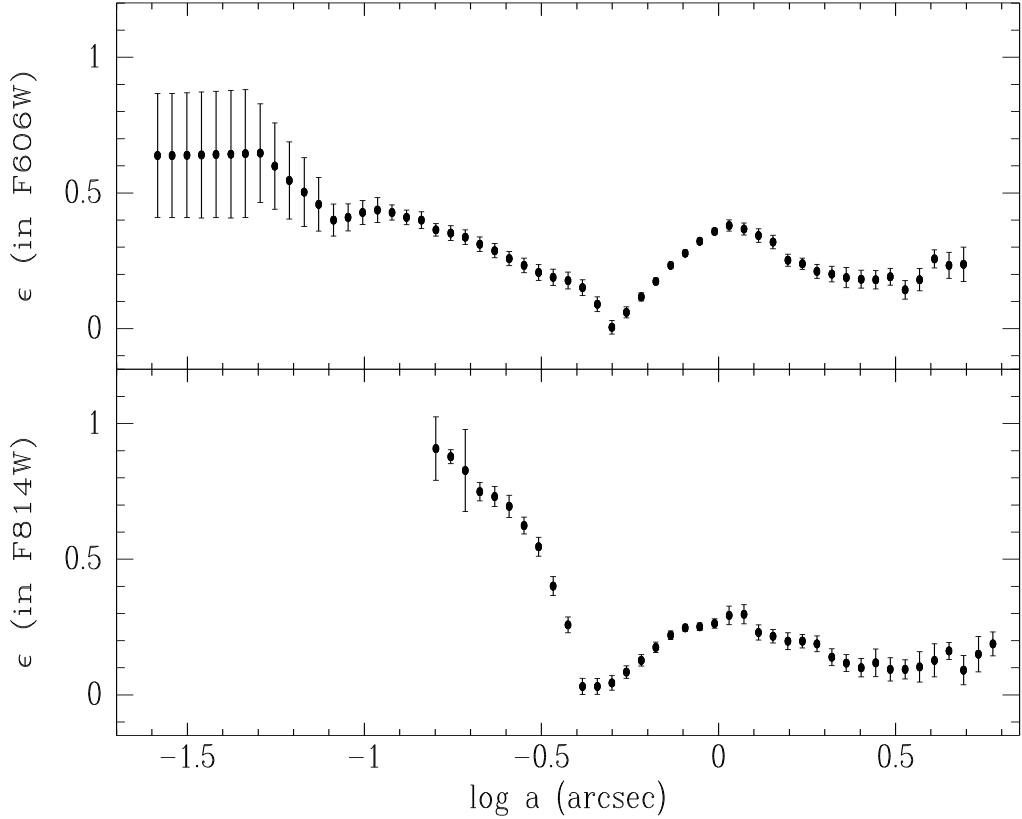}%
\caption{\label{Fig:maps}Ellipticity as a function of the semimajor axis in the F606W and F814W
filters of ACS/{\sl HST}.}
\end{figure}

\begin{figure}[!hb]
\centering
\includegraphics[height=150mm,angle=-90]{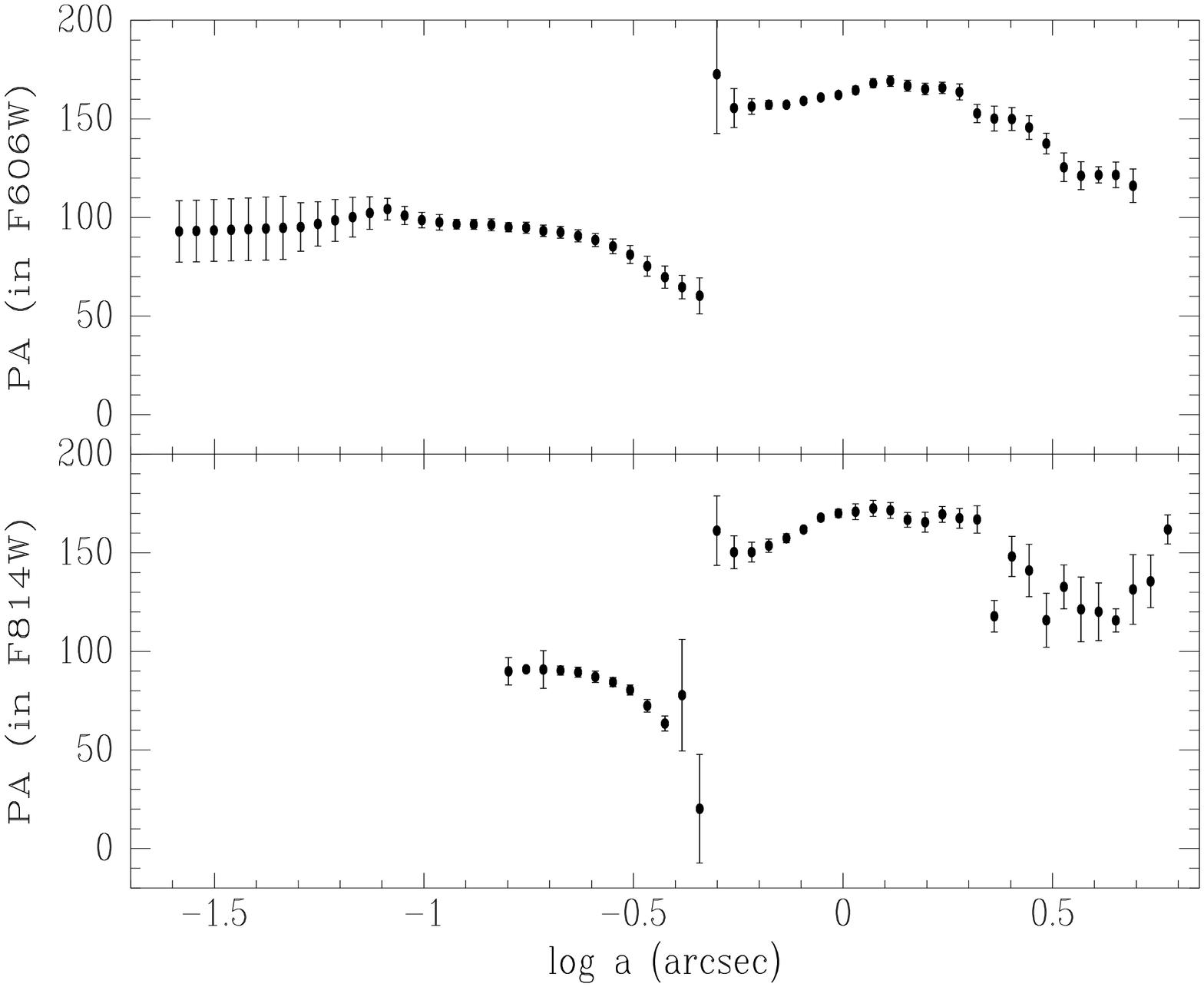}%
\caption{\label{Fig:maps}P.A. as a function of the semimajor axis in the F606W and F814W filters
of ACS/{\sl HST}.}
\end{figure}

\subsection{Point spread function}

At a distance of 780 kpc, the ACS/WFC has a scale of $\rm {0.05~arcsec=0.19~pc~pixel^{-1}}$, and
thus M31 clusters are clearly resolved with it. Their observed core structures, however, are still
affected by the PSF. We chose not to deconvolve the data, instead fitting structural models after
convolving them with a simple analytic description of the PSF as \citet{barmby07} did.
To estimate the PSF for the WFC,
\citet{barmby07} used the {\sc iraf} task {\sc ellipse}
with circular symmetry enforced to produce intensity profiles out
to radii of about $2''$ (40 pixels) for a number of isolated stars on a
number of images, and combined them to produce a single, average
PSF. This was done separately for the F606W and F814W
filters. They originally tried to fit these with simple Moffat profiles
(with backgrounds added), but found that a better description was
given by a function of the form below. For the
combination of the WFC and F606W filter,
\begin{equation}
I_{\rm PSF}=I_0\left[1+\left(R/0\farcs0686\right)^3\right]^{-1.23}\ ,
\end{equation}
which has a full width at half-maximum of ${\rm FWHM}=0\farcs125$, or about 2.5 px; for the
combination of the WFC and F814W filter,
\begin{equation}
I_{\rm PSF}=I_0\left[1+\left(R/0\farcs0783\right)^3\right]^{-1.19}\ ,
\end{equation}
which has a full width at half-maximum of ${\rm FWHM}=0\farcs145$, or about 2.9 px. In addition,
since this PSF formula is radially symmetric and the models of \citet{king66} we fit are
intrinsically spherical, the convolved models to be fitted to the data are also circularly
symmetric.

\subsection{Extinction}

When we fit models to the brightness profiles of B037, we will correct the inferred magnitude
parameters for extinction. The reddening law from \citet{car89} is employed in this paper. The
effective wavelengths of the ACS F606W and F814W filters are $\lambda_{\rm eff}=5918$ and 8060
\AA~\citep{sirianni05}, so that from \citet{car89}, $A_{\rm {F606W}}\simeq 2.8\times E(B-V)$ and
$A_{\rm {F814W}}\simeq 1.8\times E(B-V)$ \citep[see][for details]{barmby07,mclaughlin08}. The
reddening value of $E(B-V)=1.360\pm 0.013$ from \citet{ma06b} is adopted in this paper.

\subsection{Magnitudes of B037 in F606W and F814W filters}

We derived the total flux of B037 in F606W and F814W filter bands using the {\sc iraf} task {\sc
phot} in {\sc dapphot} as below: measuring aperture magnitudes in concentric apertures with an
interval of $0.1\arcsec$, drawing magnitude growth curves, and paying attention to where the flux
does not increase. At last, we obtained the magnitudes of B037 in F606W and F814W to be
$16.21\pm0.010$ and $14.16\pm0.006$, respectively. In the photometry, we derived the background
value as the mean of a region far away from the cluster (see \S 2.3 for details). We use VEGAMAG
photometric system. In order to allow a meaningful comparison with the previous ground-based
broad-band photometry of \citet{bh00}, we transformed the magnitudes from the ACS system to the
standard broad-band photometric system by following the transformation equations and coefficients
of Table 22 of \citet{sirianni05}. The results are $m_V {(\rm ACS)}=16.83$ (this paper) versus
$m_V=16.82$ \citep{bh00}, and $m_I {(\rm ACS)}=14.15$ (this paper) versus $m_I=14.16$
\citep{bh00}. Our results are in good agreement with \citet{bh00}.

\section{Models and Fits}

\subsection{Structural models}

After elliptical galaxies, GCs are the best understood and most thoroughly modelled class of
stellar systems. For example, a large majority of the $\sim 150$ Galactic GCs have been fitted by
the simple models of single-mass, isotropic, lowered isothermal spheres developed by
\citet{michie63} and \citet{king66} (hereafter ``King models''), yielding comprehensive catalogs
of cluster structural parameters and physical properties \citep[see][and references
therein]{McLaughlin05}. For extragalactic GCs, {\sl HST} imaging data have been used to fit King
models to a large number of GCs in M31 \citep[e.g.,][and references therein]{bk02b,barmby07}, in
M33 \citep{Larsen02}, and in NGC 5128 \citep[e.g.,][and references
therein]{harris02,mclaughlin08}. In this paper, we fit the usual King models to the density
profile of B037 observed with ACS/WFC.

\subsection{Observed data}

Tables 4 and 5 list the surface brightness, $\mu$, of B037, and its integrated magnitude, $m$, as
a function of radius in the F606W and F814W filters, respectively. The errors in the surface
brightness were also generated by the {\sc iraf} task {\sc ellipse}, in which they are obtained
directly from the root mean square scatter of the intensity data along the zero-ellipticity
isophotes. In addition, the surface photometries at radii where the ellipticity and position angle
cannot be measured, are obtained based on the last ellipticity and position angle as the {\sc
iraf} task {\sc ellipse} is designed.

\begin{table}
\begin{center}
\renewcommand\arraystretch{1}
\caption{B037: Surface brightness, $\mu$, and integrated magnitude, $m$, as a function of the
radius in the F606W filter of {\sl HST} ACS-WFC}
\begin{tabular}{ccc|ccc}
\hline\hline
   $R$    &  $\mu$  &   $m$   &   $R$    &  $\mu$  &   $m$  \\
 (arcsec) &  (mag)  &  (mag)  & (arcsec) &  (mag)  &  (mag) \\
\hline \hline
 0.0260  & $17.327  \pm0.007   $ & 23.827   & 0.3757  & $17.792  \pm0.040   $ & 18.456  \\
 0.0287  & $17.328  \pm0.008   $ & 23.827   & 0.4132  & $17.863  \pm0.046   $ & 18.264  \\
 0.0315  & $17.328  \pm0.008   $ & 23.827   & 0.4545  & $17.944  \pm0.049   $ & 18.123  \\
 0.0347  & $17.329  \pm0.009   $ & 23.827   & 0.5000  & $18.040  \pm0.049   $ & 17.962  \\
 0.0381  & $17.330  \pm0.010   $ & 23.827   & 0.5500  & $18.148  \pm0.048   $ & 17.827  \\
 0.0420  & $17.331  \pm0.011   $ & 23.827   & 0.6050  & $18.267  \pm0.048   $ & 17.672  \\
 0.0461  & $17.331  \pm0.012   $ & 23.827   & 0.6655  & $18.389  \pm0.053   $ & 17.549  \\
 0.0508  & $17.332  \pm0.014   $ & 22.086   & 0.7321  & $18.495  \pm0.061   $ & 17.414  \\
 0.0558  & $17.334  \pm0.015   $ & 22.086   & 0.8053  & $18.598  \pm0.070   $ & 17.295  \\
 0.0614  & $17.338  \pm0.016   $ & 22.086   & 0.8858  & $18.716  \pm0.077   $ & 17.165  \\
 0.0676  & $17.341  \pm0.018   $ & 22.086   & 0.9744  & $18.854  \pm0.079   $ & 17.039  \\
 0.0743  & $17.346  \pm0.020   $ & 21.452   & 1.0718  & $19.006  \pm0.077   $ & 16.927  \\
 0.0818  & $17.351  \pm0.022   $ & 21.452   & 1.1790  & $19.193  \pm0.107   $ & 16.822  \\
 0.0899  & $17.356  \pm0.024   $ & 21.452   & 1.2969  & $19.440  \pm0.105   $ & 16.731  \\
 0.0989  & $17.363  \pm0.027   $ & 21.452   & 1.4266  & $19.721  \pm0.116   $ & 16.642  \\
 0.1088  & $17.372  \pm0.027   $ & 21.062   & 1.5692  & $20.001  \pm0.113   $ & 16.570  \\
 0.1197  & $17.382  \pm0.029   $ & 20.552   & 1.7261  & $20.293  \pm0.116   $ & 16.505  \\
 0.1317  & $17.395  \pm0.028   $ & 20.552   & 1.8987  & $20.597  \pm0.122   $ & 16.451  \\
 0.1448  & $17.411  \pm0.028   $ & 20.370   & 2.0886  & $20.872  \pm0.128   $ & 16.401  \\
 0.1593  & $17.429  \pm0.029   $ & 19.964   & 2.2975  & $21.224  \pm0.105   $ & 16.358  \\
 0.1752  & $17.450  \pm0.028   $ & 19.964   & 2.5272  & $21.457  \pm0.112   $ & 16.320  \\
 0.1928  & $17.475  \pm0.026   $ & 19.764   & 2.7800  & $21.719  \pm0.138   $ & 16.283  \\
 0.2120  & $17.504  \pm0.026   $ & 19.528   & 3.0580  & $22.082  \pm0.154   $ & 16.251  \\
 0.2333  & $17.538  \pm0.025   $ & 19.337   & 3.3638  & $22.603  \pm0.164   $ & 16.225  \\
 0.2566  & $17.578  \pm0.026   $ & 19.091   & 3.7001  & $23.042  \pm0.225   $ & 16.206  \\
 0.2822  & $17.624  \pm0.026   $ & 19.009   & 4.0701  & $23.694  \pm0.467   $ & 16.191  \\
 0.3105  & $17.675  \pm0.030   $ & 18.802   & 4.4772  & $24.509  \pm0.571   $ & 16.182  \\
 0.3415  & $17.732  \pm0.034   $ & 18.634   & 4.9249  & $25.173  \pm1.342   $ & 16.172  \\
 \hline \hline
\end{tabular}
\end{center}
\end{table}

\begin{table}
\begin{center}
\renewcommand\arraystretch{1}
\caption{B037: Surface brightness, $\mu$, and integrated magnitude, $m$, as a function of the
radius in the F814W filter of {\sl HST} ACS-WFC}
\begin{tabular}{ccc|ccc}
\hline\hline
   $R$    &  $\mu$  &   $m$   &   $R$    &  $\mu$  &   $m$  \\
 (arcsec) &  (mag)  &  (mag)  & (arcsec) &  (mag)  &  (mag) \\
\hline \hline
 0.0260  & $15.301  \pm0.010   $ & 21.800   & 0.4132  & $15.772  \pm0.032   $ & 16.190  \\
 0.0287  & $15.302  \pm0.011   $ & 21.800   & 0.4545  & $15.863  \pm0.036   $ & 16.048  \\
 0.0315  & $15.303  \pm0.012   $ & 21.800   & 0.5000  & $15.967  \pm0.038   $ & 15.889  \\
 0.0347  & $15.303  \pm0.013   $ & 21.800   & 0.5500  & $16.078  \pm0.037   $ & 15.754  \\
 0.0381  & $15.304  \pm0.015   $ & 21.800   & 0.6050  & $16.190  \pm0.033   $ & 15.598  \\
 0.0420  & $15.305  \pm0.016   $ & 21.800   & 0.6655  & $16.300  \pm0.036   $ & 15.474  \\
 0.0461  & $15.306  \pm0.018   $ & 21.800   & 0.7321  & $16.407  \pm0.046   $ & 15.338  \\
 0.0508  & $15.308  \pm0.019   $ & 20.061   & 0.8053  & $16.543  \pm0.053   $ & 15.218  \\
 0.0558  & $15.310  \pm0.021   $ & 20.061   & 0.8858  & $16.706  \pm0.054   $ & 15.094  \\
 0.0614  & $15.313  \pm0.024   $ & 20.061   & 0.9744  & $16.847  \pm0.054   $ & 14.974  \\
 0.0676  & $15.317  \pm0.026   $ & 20.061   & 1.0718  & $16.995  \pm0.061   $ & 14.868  \\
 0.0743  & $15.321  \pm0.030   $ & 19.430   & 1.1790  & $17.185  \pm0.105   $ & 14.767  \\
 0.0818  & $15.326  \pm0.033   $ & 19.430   & 1.2969  & $17.459  \pm0.086   $ & 14.681  \\
 0.0899  & $15.331  \pm0.037   $ & 19.430   & 1.4266  & $17.717  \pm0.107   $ & 14.596  \\
 0.0989  & $15.337  \pm0.041   $ & 19.430   & 1.5692  & $18.006  \pm0.101   $ & 14.527  \\
 0.1088  & $15.347  \pm0.043   $ & 19.039   & 1.7261  & $18.279  \pm0.102   $ & 14.464  \\
 0.1197  & $15.360  \pm0.043   $ & 18.531   & 1.8987  & $18.567  \pm0.086   $ & 14.411  \\
 0.1317  & $15.367  \pm0.045   $ & 18.531   & 2.0886  & $18.833  \pm0.092   $ & 14.359  \\
 0.1448  & $15.375  \pm0.048   $ & 18.350   & 2.2975  & $19.167  \pm0.095   $ & 14.316  \\
 0.1593  & $15.396  \pm0.046   $ & 17.940   & 2.5272  & $19.460  \pm0.094   $ & 14.277  \\
 0.1752  & $15.411  \pm0.046   $ & 17.940   & 2.7800  & $19.649  \pm0.156   $ & 14.240  \\
 0.1928  & $15.428  \pm0.045   $ & 17.738   & 3.0580  & $20.075  \pm0.137   $ & 14.207  \\
 0.2120  & $15.447  \pm0.046   $ & 17.496   & 3.3638  & $20.538  \pm0.103   $ & 14.181  \\
 0.2333  & $15.466  \pm0.048   $ & 17.301   & 3.7001  & $21.002  \pm0.138   $ & 14.160  \\
 0.2566  & $15.501  \pm0.046   $ & 17.046   & 4.0701  & $21.399  \pm0.203   $ & 14.142  \\
 0.2822  & $15.531  \pm0.043   $ & 16.960   & 4.4772  & $21.964  \pm0.228   $ & 14.129  \\
 0.3105  & $15.580  \pm0.037   $ & 16.745   & 4.9249  & $22.519  \pm0.240   $ & 14.117  \\
 0.3415  & $15.632  \pm0.031   $ & 16.573   & 5.4174  & $23.311  \pm0.588   $ & 14.106  \\
 0.3757  & $15.695  \pm0.029   $ & 16.388   & 5.9591  & $23.508  \pm0.782   $ & 14.096  \\
\hline \hline
\end{tabular}
\end{center}
\end{table}

\subsection{Fits}

Our fitting procedure involves computing in full large numbers of King structural models, spanning
a wide range of fixed values of the appropriate shape parameter $W_0$ \citep[see][in
detail]{McLaughlin05}. And then the models are convolved with the ACS/WFC PSF for the F606W and
F814W filters of equations of (1) and (2):

\begin{equation}
\widetilde{I}_{\rm mod}^{*} (R | r_0) = \int\!\!\!\int_{-\infty}^{\infty}
               \widetilde{I}_{\rm mod}(R^\prime/r_0) \times
               \widetilde{I}_{\rm PSF}
                    \left[(x-x^\prime),(y-y^\prime)\right]
       \ dx^\prime \, dy^\prime\ ,
\label{eq:convol}
\end{equation}
where $\widetilde{I}_{\rm mod}\equiv I_{\rm mod}/I_0$; and $\widetilde{I}_{\rm PSF}$
is the PSF profile
normalized to unit total luminosity \citep[see][in detail]{mclaughlin08}. We
changed the luminosity density to surface brightness $\widetilde{\mu}_{\rm
mod}^{*}=-2.5\,\log\,[\widetilde{I}_{\rm mod}^{*}]$ before fitting them to the observed
surface-brightness profile of B037, $\mu=\mu_{0}-2.5\,\log\,[I(R/r_0)/I_0]$, finding the radial
scale $r_0$ and central surface brightness $\mu_{0}$ which minimize $\chi^2$ for every given value
of $W_0$. The $(W_0,r_0,\mu_{0})$ combination that yields the global minimum $\chi_{\rm min}^2$
over the grid used defines the best-fit model of that type:
\begin{equation}
\chi^2  =
  \sum_i{
 \frac{\left[\mu_{\rm obs}(R_i)
             - \widetilde{\mu}_{\rm mod}^{*}(R_i | r_0)
             \right]^2}
      {\sigma_i^2}
        } ,
\label{eq:chi2}
\end{equation}
in which $\sigma_i$ is the error in the surface brightness. Estimates of the one-sigma
uncertainties on these basic fit parameters follow from their extreme values over the subgrid of
fits with $\chi^2/\nu\le \chi_{\rm min}^2/\nu+1$, here $\nu$ is the number of free parameters.
Figure 5 shows our best King fits to B037. In Fig. 5, open squares are {\sc ellipse} data points
included in the least-squares model fitting, and the asterisks are points not used to constrain
the fit. These observed data points shown by asterisks are included in the radius of
$R<2~\rm{pixels}=0\farcs1$, and the isophotal intensity is dependent on its neighbors. As
\citet{barmby07} pointed out that, the {\sc ellipse} output contains brightnesses for 15 radii
inside 2 pixel, but they are all measured from the same 13 central pixels and are not
statistically independent. So, to avoid excessive weighting of the central regions of B037 in the
fits, we only used intensities at radii $R_{\rm min}$, $R_{\rm min}+(0.5,1.0,2.0~{\rm pixels})$,
or $R>2.5~{\rm pixels}$ as \citet{barmby07} used. Table 6 summarizes the results obtained in this paper.

\begin{figure}
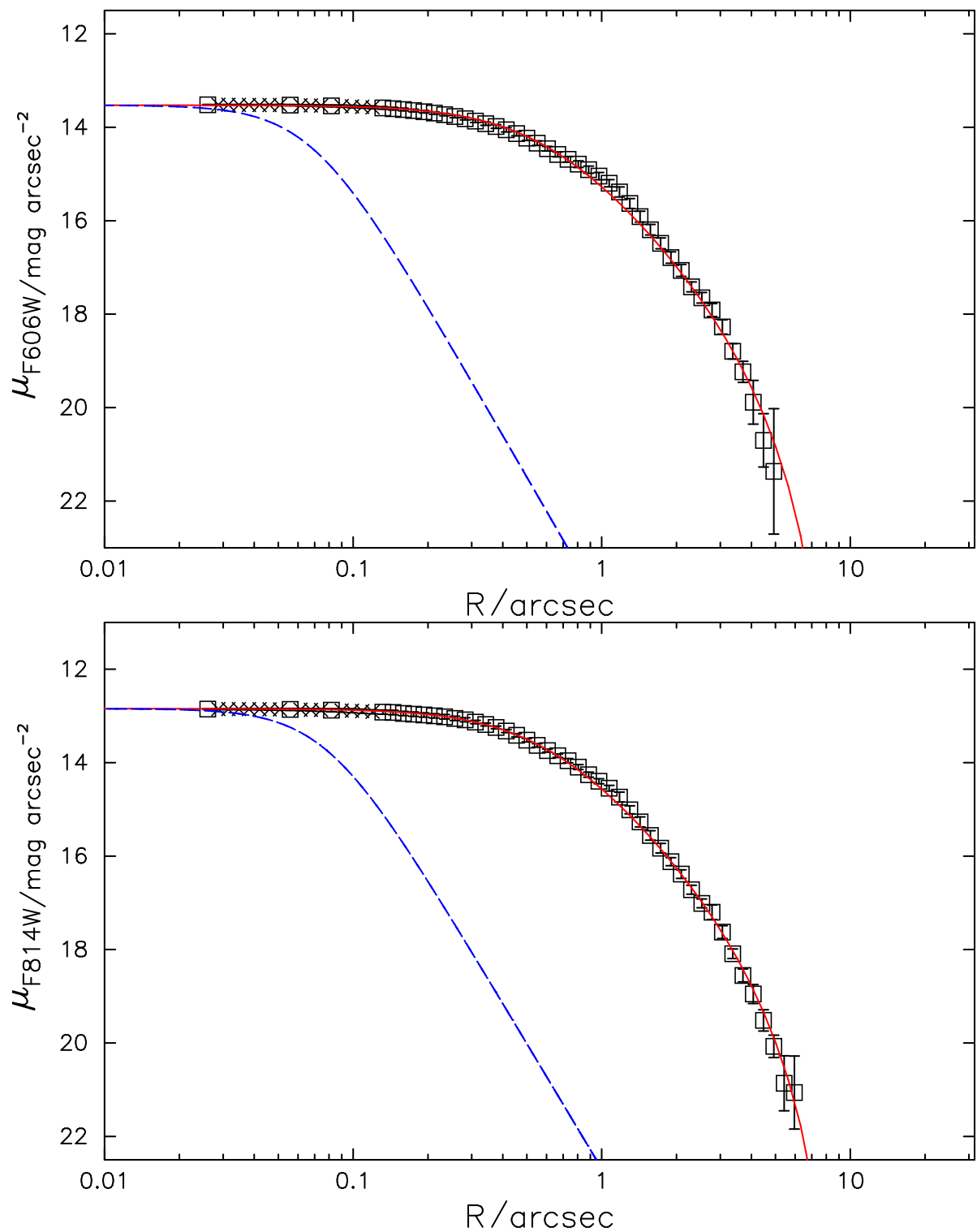

\centering
\includegraphics[height=100mm,angle=0]{ms602fig5a.eps}\\
\includegraphics[height=100mm,angle=0]{ms602fig5b.eps}%
\caption{\label{Fig:spectra}Surface brightness profile of B037 measured in the F606W and F814
filters. Dashed curve (blue) trace the PSF intensity profiles and solid (red) curves are the
PSF-convolved best-fit models. Open squares are {\sc ellips} data points included in the $\chi^2$
model fitting, and the asterisks are points not used to constrain the fits (see the text in
detail).}
\end{figure}

\begin{table}
\begin{center}
\renewcommand\arraystretch{1}
\caption{\label{Tab:assocation} Structural parameters of B037}
\begin{tabular}{lcc}
\hline \hline
Parameters &             F606W                            &       F814W   \\
\hline
$r_0$      & $0.56\pm0.02\arcsec~(=2.16\pm0.08~\rm{pc})$  &
$0.56\pm0.01\arcsec~(=2.16\pm0.04~\rm{pc})$ \\
$r_t$      & $8.6\pm0.4\arcsec~(=33.1\pm1.5~\rm{pc})$     &
$8.9\pm0.3\arcsec~(=34.3\pm1.2~\rm{pc})$    \\
$c=\log (r_t/r_0)$ &        $1.19\pm0.02$                 &   $1.20\pm0.01$ \\
$r_h$      &  $1.05\pm0.03\arcsec(=4.04\pm0.12~\rm{pc})$    &
$1.07\pm0.01\arcsec(=4.12\pm0.04~\rm{pc})$    \\
$\mu_0$ (${\rm mag~arcsec^{-2}}$)    &  $13.53\pm 0.03$               &   $12.85\pm 0.03$ \\
\hline \hline
\end{tabular}
\end{center}
\end{table}

\subsection{Comparison to previous results}

\citet{ma06a} analyzed the same F606W image of B037 used here, fitting a \citet{king62} model to a
surface brightness profile derived from a PSF-deconvolved image. They derived the scale radius
$r_0=0\farcs72$ (it is called the core radius in \citet{ma06a}), half-light radius
$r_h=1\farcs11$, concentration index $c=0.91$, and central surface brightness $\mu(0)=17.21~{\rm
mag~arcsec^{-2}}$ (using the value for extinction adopted in this paper, this becomes
$\mu_0=13.40~{\rm mag~arcsec^{-2}}$). Comparing the results of \citet{ma06a} with Table 6 of this
paper, we find that our model fits produce a somewhat higher concentration and smaller scale
radius. These differences come from: (i) using different models (\citet{king62} vs
\citet{king66}), (ii) the observed data are obtained with different ways. In (ii), \citet{ma06a}
derived the surface brightness profile from a PSF-deconvolved image; in addition, \citet{ma06a}
derived the surface brightness profile with ellipticity and position angle allowed to vary with
the isophote semimajor axis, however, in this paper, we derived the surface brightness profile on
fixed, zero-ellipticity isophotes, since we choose to fit circular models for the intrinsic
cluster structure and the PSF as \citet{barmby07} did (see \S 2.4 for details). In fact, from Fig.
5 of this paper and Fig. 3 of \citet{ma06a}, we can see that, the observed data are somewhat
different between \citet{ma06a} and this paper.

\citet{barmby07} analyzed the same F606W and F814W images of B037 used here with nearly the same
observed data and method. The results of comparison are listed in Table 7 (Table 5 of
\citet{barmby07} in the electronic edition did not list the results of B037 in the F814W filter.),
from which we can see that the results obtained in this paper are in good agreement with ones of
\citet{barmby07} (about the central surface brightnesses, we have corrected them using the value
for extinction adopted in this paper).

\begin{table}
\begin{center}
\renewcommand\arraystretch{1}
\caption{\label{Tab:assocation} Results of comparison}
\begin{tabular}{lcc|cc}
\hline \hline \multicolumn{1}{l}{} & \multicolumn{2}{c|}{F606W} &
\multicolumn{2}{c}{F814W} \\
\hline
Parameters & \citet{barmby07}  &  This paper  &  \citet{barmby07}  &  This paper \\
\hline $r_0$       &  0\farcs56   &  0\farcs56  & 0\farcs59 &   0\farcs56 \\
$c=\log (r_t/r_0)$ &  1.23   &  1.19  & 1.18 &   1.20 \\
$r_h$              &  1\farcs09   &  1\farcs05  & &   1\farcs07 \\
$\mu_0$ (${\rm mag~arcsec^{-2}}$) & 13.45 & 13.53  &   12.75          &   12.85 \\
\hline \hline
\end{tabular}
\end{center}
\end{table}

\section{Discussion and summary}

As discussed in \S 3.1, it is impossible that the dust lane comes from the Milky Way. Another
possibility is that the dust lane is contained in B037 itself. As we know, the formation of dust
requires gas with a rather high metallicity. \citet{perr02} presented metallicities for more than
200 GCs in M31 including B037, using the Wide Field Fibre Optic Spectrograph at the 4.2 m William
Herschel Telescope in La Palma, Canary Islands, which provides a total spectral coverage of $\sim$
3700-5600~\rm \AA~ with two gratings. One grating (H2400B 2400 line) yielded a dispersion of
0.8~\rm\AA ${\rm~pixel^{-1}}$ and  a spectral resolution of 2.5~\rm\AA~ over the range
3700-4500~\rm\AA~ covering the CN feature at 3883~\AA, the H and K lines of calcium, $\rm
H\delta$, the CH~G band and the 4000~\AA~ continuum break, and the other grating (R1200R 1200
line) presented a dispersion of 1.5~\rm\AA${\rm~pixel^{-1}}$ and a spectral resolution of
5.1~\rm\AA~ over the range 4400-5600~\rm\AA~ to add absorption features such as $\rm H\beta$, the
Mg~$b$ triplet, and two iron lines near 5300~\AA. Then, \citet{perr02} calculated 12
absorption-line indices based on the prescription of \citet{bh90}. By the comparison of the line
indices with the published M31 GC [Fe/H] values from the previous literature
\citep{bonoli87,bh90,bh00}, the results of linear fits were obtained. Final cluster metallicities
were determined from an unweighted mean of the [Fe/H] values calculated from the CH (G), Mg~$b$,
and Fe53 line strengths. For B037, \citet{perr02} obtained its metallicity to be $\rm
[Fe/H]=-1.07\pm 0.20$. It is clear that B037 has a low metallicity. So, it is intricate that where
is the dust lane from?

In this paper, using the images of deep observations and of highly spatial resolutions with the
ACS/{\sl HST}, we firstly present that the GC B037 in M31 is crossed by a dust lane. Photometric
data in the F606W and F814W bands provide that, colors of ($\rm {F606W-F814W}$) in the dust lane
are redder $\sim 0.4$ mags than ones in the other regions of B037. From the {\sl HST} images, this
dust lane seems to be contained in B037, not from the Milky Way. However, the formation of dust
requires gas with a rather high metallicity. So, it seems impossible that the observed dust lane
is physically associated with B037 itself, which has a low metallicity to be $\rm [Fe/H]=-1.07\pm
0.20$ from \citet{perr02}. So, that the observed dust lane in the view of B037 is from B037 itself
or from the Milky Way needs to be confirmed in the future. In addition, based on these images, we
present the precise variation of ellipticity and position angle, and of surface brightness
profile, and determine the structural parameters of B037 by fitting a single-mass isotropic King
model.

\begin{acknowledgements}
I am indebted to Daming Chen, Zhou Fan, Tianmeng Zhang and Song Wang for their helps in finishing
this paper. I am also grateful to the referee for the important comments.
This work was supported by the Chinese National Natural Science Foundation grands No.
10873016, and 10633020, and by National Basic Research Program of China (973 Program), No.
2007CB815403.
\end{acknowledgements}


\begin{thebibliography}{}
\bibitem[Barmby et al.(2002a)]{bk02b}Barmby, P., Holland, S., \&
Huchra, J. 2002a, \aj, 123, 1937

\bibitem[Barmby et al.(2000)]{bh00}Barmby, P., Huchra, J., Brodie, J.,
Forbes, D., Schroder, L., \& Grillmair, C. 2000, \aj, 119, 727

\bibitem[Barmby et al. (2007)]{barmby07}Barmby P., McLaughlin D. E.,
Harris W. E., Harris G. L. H., \& Forbes D. A. 2007, AJ, 133, 2764

\bibitem[Barmby et al.(2002b)]{bk02}Barmby, P., Perrett, K. M., \& Bridges, T.
J. 2002b, MNRAS, 329, 461




\bibitem[Battistini et al.(1987)]{battis87}Battistini, P.,
B\`{o}noli, F., Braccesi, A., Federici, L., Fusi Pecci F., Marano, B., \& B\"{o}rngen, F. 1987,
\aaps, 67, 447

\bibitem[B\`{o}noli et al.(1987)]{bonoli87}B\`{o}noli, F., Delpino, F.,
Federici, L., \& Fusi Pecci, F. 1987, A\&A, 185, 25

\bibitem[Brodie \& Huchra(1990)]{bh90} Brodie, J. P., \& Huchra,
J. P. 1990, ApJ, 362, 503



\bibitem[Cardelli et al.(1989)]{car89}Cardelli, J. A.,
Clayton, G. C., \& Mathis, J. S. 1989, \apj, 345, 245

\bibitem[Crampton et al.(1985)]{cram85}Crampton, D., Cowley, A. P.,
Schade, D., \& Chayer, P. 1985, \apj, 288, 494


\bibitem[Fan et al.(2008)]{fan08}Fan, Z., Ma, J., de Grijs, R., \&
Zhou, X. 2008, MNRAS, 385, 1973



\bibitem[Galleti et al.(2004)]{gall04}Galleti, S., Federici, L.,
Bellazzini, M., Fusi Pecci, F., \& Macrina, S. 2004, A\&A, 416, 917


\bibitem[Galleti et al.(2006)]{gall06}Galleti, S., Federici, L.,
Bellazzini, M., Buzzoni, A., \& Fusi Pecci, F. 2006, A\&A, 456, 985

\bibitem[Galleti et al.(2007)]{gall07}Galleti, S., Bellazzini, M.,
Federici, L., Buzzoni, A., \& Fusi Pecci, F. 2007, A\&A, 471, 127

\bibitem[Harris et al.(2002)]{harris02}Harris W. E., Harris G. L. H.,
Holland S. T., \& McLaughlin D. E. 2002, AJ, 124, 1435


\bibitem[Hubble(1932)]{hubble32}Hubble, E. P. 1932, \apj, 76, 44

\bibitem[Huchra et al.(1991)]{huchra91}Huchra, J. P., Brodie, J.
P., \& Kent, S. M. 1991, \apj, 370, 495

\bibitem[King(1962)]{king62}King, I. R. 1962, \aj, 67, 471

\bibitem[King(1966)]{king66}King, I. R. 1966, AJ, 71, 64

\bibitem[Kron \& Mayall(1960)]{kronmay60}Kron, G. E., \& Mayall, N. U.
1960, \aj, 65, 581

\bibitem[Larsen et al.(2002)]{Larsen02}Larsen, S. S., Brodie, J. P.,
Sarajedini, A., \& Huchra, J. P. 2002, AJ, 124, 2615



\bibitem[Ma et al.(2007)]{ma07}Ma, J., et al. 2007, MNRAS, 376, 1621

\bibitem[Ma et al.(2006a)]{ma06b}Ma, J., de Grijs, R., Yang, Y., Zhou, X.,
Chen, J., Jiang, Z., Wu, Z., \& Wu, J. 2006a, MNRAS, 368, 1143

\bibitem[Ma et al.(2006b)]{ma06a}Ma, J., van den Bergh,
S., \& Wu, H., et al. 2006b, \apj, 636, L93

\bibitem[Mackey \& van den Bergh(2005)]{mackey05}Mackey, A., \& van den Bergh,
S. 2005, \mnras, 360, 631

\bibitem[McLaughlin et al.(2008)]{mclaughlin08}
McLaughlin, D. E., Barmby, P., Harris, W. E., Forbes, D. A., \& Harris, G. L. H. 2008, \mnras,
384, 563

\bibitem[McLaughlin \& van der Marel(2005)]{McLaughlin05}
McLaughlin, D. E., \& van der Marel, R. P. 2005, ApJS, 161, 304

\bibitem[Meylan \& Heggie(1997)]{meylan97}Meylan, G., \& Heggie,
D. 1997, A\&A Rev., 8, 1

\bibitem[Meylan et al.(2001)]{meylan01}Meylan, G., Sarajedini, A.,
Jablonka, P., Djorgovski, S., Bridges, T., \&  Rich, R. 2001, \aj, 122, 830

\bibitem[Michie(1963)]{michie63}Michie, R. W. 1963, MNRAS, 125, 127

\bibitem[Perrett et al.(2002)]{perr02}Perrett, K. M.,
et al. 2002, \aj, 123, 2490


\bibitem[Sirianni et al.(2005)]{sirianni05}Sirianni, M., et al. 2005, \pasp, 117, 1049


\bibitem[Stetson(1987)]{stet87}Stetson, P. B. 1987, \pasp, 99, 191

\bibitem[van den Bergh(1968)]{Sidney68}van den Bergh, S. 1968,
The Observatory, 88, 168




\bibitem[Vete\u{s}nik(1962a)]{vete62a}Vete\u{s}nik, M. 1962a,
BAC, 13, no. 5, p. 180

\bibitem[Vete\u{s}nik(1962b)]{vete62b}Vete\u{s}nik, M. 1962b,
BAC, 13, no. 6, p. 218



\end{thebibliography}
\end{document}